# Modern High Intensity H- Accelerator Sources

*Vadim Dudnikov, Muons, Inc, Batavia,*
*IL USA*

Vadim@muonsinc.com

*Abstract*— A review of modern high intensity H- ion sources for accelerators is presented. The cesiation effect, a significant enhancement of negative ion emission from gas discharges with decrease of co-extracted electron current below negative ion current, was observed for the first time by inserting into the discharge chamber a compound with one milligram of cesium on July 1, 1971 in the Institute of Nuclear Physics (INP), Novosibirsk, Russia [1]. This observation became the basis for the development of surface plasma negative ion sources (SPS) [2,3]. The efficiency of negative ion generation was increased by the invention of geometrical focusing [4]. The magnetron-planotron with geometrical focusing SPS is discussed. The converter SPS [5] is reviewed. Semiplanotron SPS [4] is discussed. Penning discharge SPS [6,7], RF pulsed and CW SPS are reviewed [8,9,10]. The history of negative ion source development is reviewed [11,12]. Large development projects, including the SPS for the Large Hadron Collider (LHC) and for the International Thermonuclear Experimental Reactor (ITER) are being conducted. The development and fabrication of injectors with cesiated SPS has become a billion dollars scale industry.

*Keywords—Surface plasma sources, negative ions, accelerators,*

## I. Introduction

Some atoms can attract by polarized forces an extra electron and form a stable negative ion with a charge -e. The stability of these negative ions is quantified by the electron affinity S, the minimum energy required to remove the extra electron. The electron affinity S is substantially smaller than the ionization energies, 0.08 eV for He- and 3.6 eV for Cl-, 0.75 eV for H-. For electron energy ~10 eV, the H- ionization cross section is ~ 30 $10^{16}$ cm$^2$, ~30 times greater than for typical neutral atoms. For H+ energies below 100 eV, the recombination cross section is greater than ~100 $10^{-16}$ cm$^2$. Charged particle collisions destroy H- ions easily.

H- ions can be formed in plasma by collision of electrons with H atoms and $H_2$ molecules.

According to energy conservation, when forming a negative ion through direct electron attachment to the atom, the excess energy has to be dissipated by photon radiation

H+e→H- +γ.

But radiative capture is rare (5 $10^{-22}$ cm$^2$ for H). More efficient a process where the excess energy can be transferred to a third particle,

$H_2$+e→ $H_2$- → H+H- ($10^{-20}$ cm$^2$ for $H_2$).

More efficient are process which excited molecules interact with low energy electrons.

$H_2 v$ +e→ H +H- (~3 $10^{-17}$cm$^2$ , for  4< v <9, e<1 eV).

The fast electrons, needed to excite the molecules, destroy the H- faster than they are produced (~3 $10^{-15}$ cm$^2$).

## II DISCOVERY OF SURFACE PLASMA NEGATIVE ION FORMATION

The development of positive and negative ion sources up to 1970 is described in the book by M. Gabovich [13]. At that time most attention was concentrated on charge exchange negative ion sources, because there was no hope to extract more than 5 mA of H-. from the plasma. The development of high brightness H$^-$ sources was first stimulated by the success of high current proton beam accumulation using charge-exchange injection [14] and then by funding from "Star Wars" [15]. This latter support led to classified work that implied difficulties and long delays of first publications, but nonofficial communication was relative fast.

In 1970 we started to work with plasma sources with planotron geometry, shown in Fig. 1.

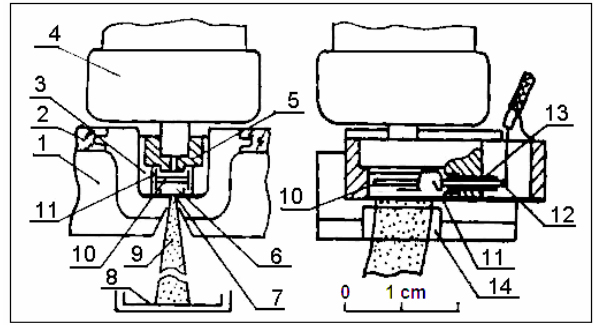

Fig. 1. Schematic of plasma source with planotron geometry.

From this ion source we can extract up to 5 mA of H-. On July 1, 1971 I inserted some cesium chromate pellets into the discharge chamber and the H- beam current increased from 1.5 mA to 15 mA [1]. Further experiments [2,3] showed that the increased H- emission was connected with the increase of secondary ion-ion emission from electrodes with decreased work function bombarded by particles from the discharge. This observation started the development of efficient surface plasma methods of negative ion production (SPM). Now the most intense beams of negative ions are produced from surface plasma sources with cesiation (SPS).

## III MAGNETRON-PLANOTRON SPS

The magnetron-planotron SPS was invented in INP, Novosibirsk [1] and was used in many accelerators. The magnetron-planotron with geometrical spherical focusing used in Brookhaven National Laboratory (BNL) is shown in Fig. 2. The hydrogen plasma produced by the 10 A arc discharge interacts with the low work function Cs-Mo cathode surface for reliable, stable operation at 100 mA peak current, 0.4 ms pulse length, ~ 0.30 mm-mrad emittance, for 6 months. It was recently tested with 1 ms pulse and duty factor of 0.73% with capability to go to 1% with present cooling. This is the highest peak current H- source used at accelerators.

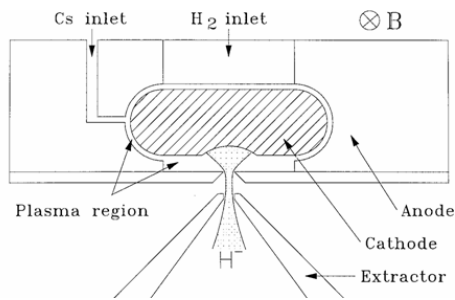

Fig. 2. Schematic of magnetron-planotron used in BNL.

A similar magnetron-planotron is used at Fermi National Accelerator Laboratory (FermiLab), Argonne IPSNS, DESY, and the China synchrotron [16].

A version with improved cooling is shown in Fig. 3 [17].

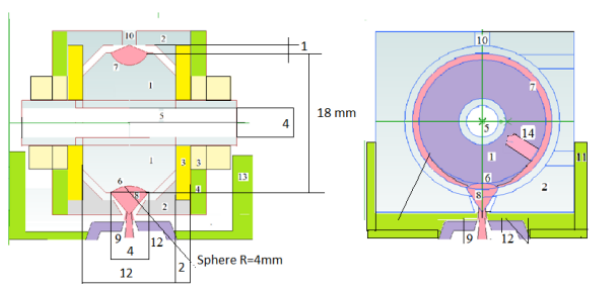

Fig. 3. Cross sections of magnetron SPS with cathode cooling. (right-median transverse to the magnetic field; left- section along the magnetic field).

For this SPS, the duty factor can be extended to 100%.

## IV SPS WITH PENNING DISCHARGE

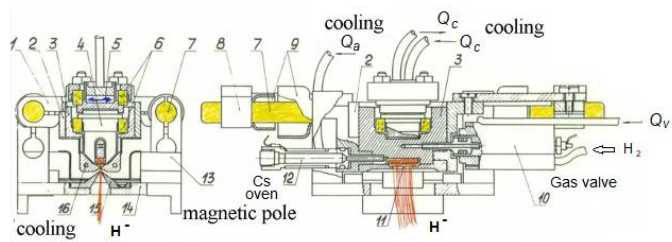

Fig. 4. Schematic diagram of SPS with a Penning discharge.

The Penning discharge SPS for production of 150 mA H- beam was developed by V. Dudnikov [18]. The schematic of this SPS is shown in Fig. 4. It was further developed in INP [19] and in Los Alamos [20]. It is used for linac injection at the Institute of Nuclear Research (INR), Moscow, Russia. Rutherford- Appleton Laboratory (RAL) started the development of SPS PD in 1979 for charge exchange injection into the ISIS synchrotron [21]. It is in operation until now, delivering 55 mA of H- beam with df 1.1% for 5 weeks [22]. The schematic of this PD SPS is shown in Fig. 5. A clone of this PD SPS was adapted for charge exchange injection in the Chinese SNS [23].

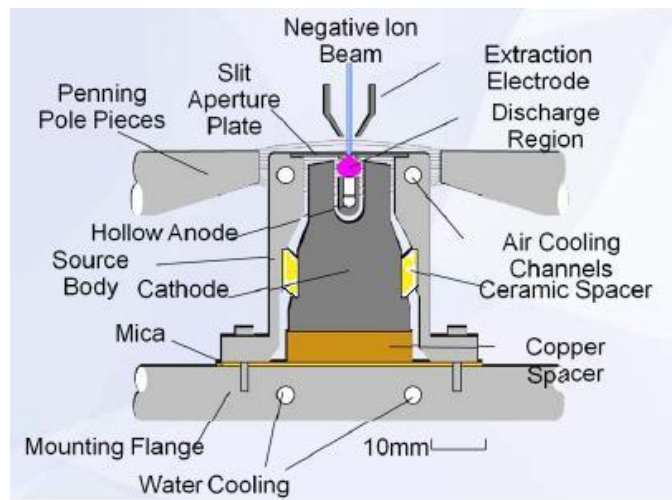

Fig. 5. Schematic of Penning discharge SPS at RAL.

## V DEVELOPMENT OF GEOMETRICAL FOCUSING

The efficiency of negative ion generation was increased significantly by the invention a geometrical focusing by V. Dudnikov [24]. The schematic of this SPS, named semiplanotron is shown in Fig. 6.

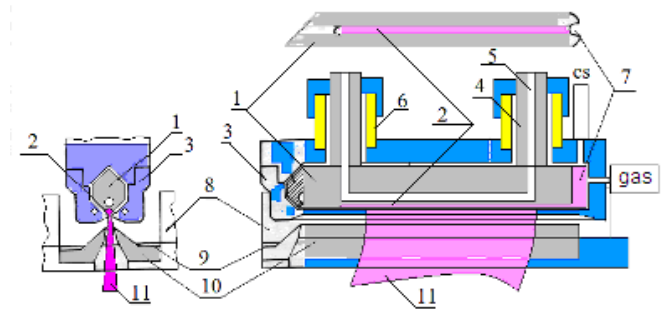

Fig. 6. Schematic of first semiplanotron with geometrical focusing.

In this SPS the discharge is glowing in the semicylindrical groove near the extraction aperture. Formed negative ions are accelerated normally to the cylindrical surface and focused to the emission slit. This SPS produced up to 0.9 A of H- with an emission slit 0.75x45 mm$^2$ with efficiency 0.3 A/kWt. A simple version of a semiplanotron for accelerators [25] was developed later as shown in Fig. 7. The discharge plasma drifts in crossed ExB fields along the groove. The positive ions accelerated by the discharge voltage bombard the groove surface and secondary negative ions emitted from this surface are accelerated by the cathode discharge voltage drop and focused to the narrow 0.5 mm slit from all cylindrical cathode surfaces The negative ion beam H- is extracted from the discharge by high voltage applied between the anode (2) and extraction electrodes (6). Interesting emission characteristics shown in Fig. 8 were observed in this semiplanotron SPS [26]. At low discharge current the emission of negative ions from the cathode surface increases with discharge current, but destruction of negative ions in the plasma reduced the emitted beam intensity exponentially after reaching the optimal value of the discharge current, and at higher discharge current it started rising again. This N-shaped emission curve was formed due to generation of negative ions on the near emission slit surface caused by a flux of fast atoms.

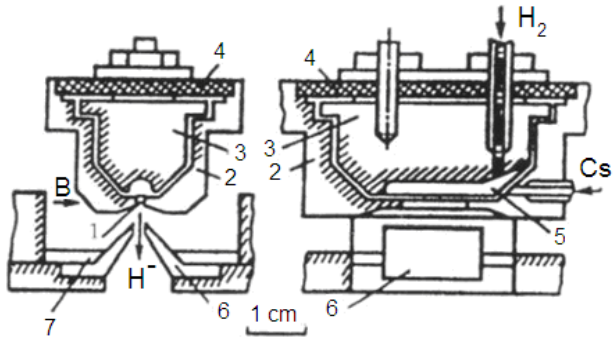

Fig. 7. Schematic of semiplanotron for accelerators.

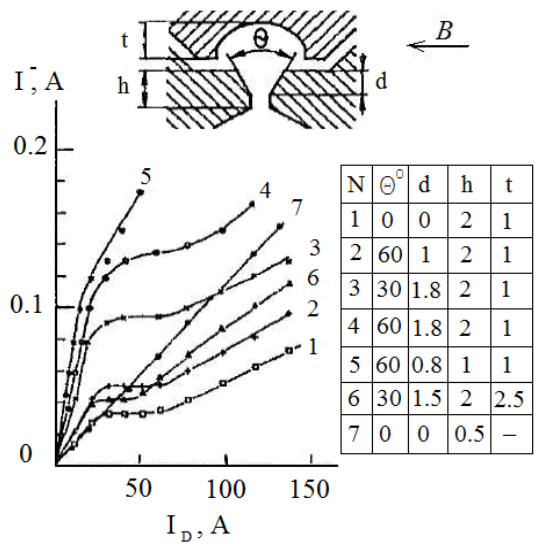

Fig. 8. Emission characteristics of semiiplanotrons with different discharge cell configurations with 0.5x10 mm$^2$ slit. (7) is for a Penning discharge SPS.

## VI CONVERTER SPS DEVELOPMENT

In 1981 Ehlers and Leung from LBNL developed a Large Volume SPS with self extraction [27] of 1 A H- beam. It has a filament discharge and negative biased converter, which focuses formed negative ions onto the emission aperture. Based on this SPS a converter SPS was developed for the Los Alamos Linac [28]. The schematic of this SPS is shown in Fig. 9. The large gas discharge chamber has diameter 17.8 cm and height 12.8 cm. Two filaments support discharge with 90 Volts. A 5 cm diameter converter biased up to -300 V emits up to 20 mA H- beam and focuses it to the 6.4 cm diameter emission aperture 8.25 cm from the converter. A similar converter SPS was developed for KEK, Japan in 1985 [29].

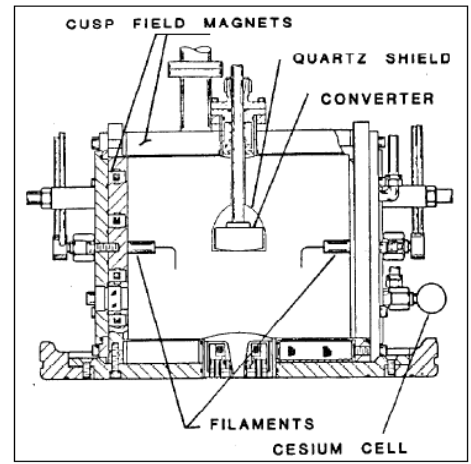

Fig. 9. Schematic of a converter SPS for Los Alamos Linac.

## VII DEVELOPMENT OF RF SPS WITH CESIATION

In 2004 LBNL developed an RF SPS with internal antenna and cesiation for the Spallation Neutron Source at ORNL [30]. The schematic of this RF SPS is shown in Fig. 10.

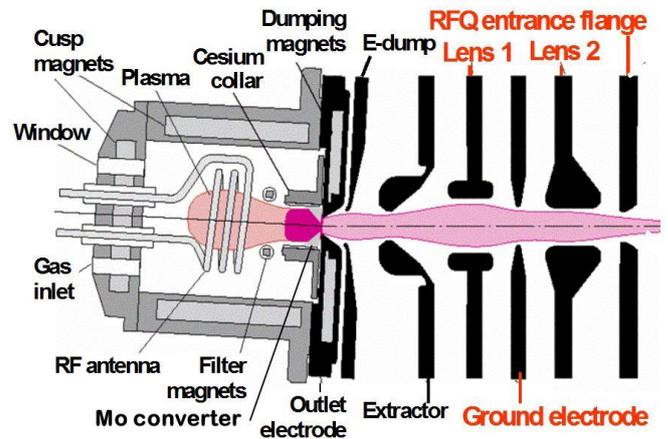

Fig. 10. Schematic of internal antenna RF SPS with cesiation in SNS ORNL.

Typically, 300 W from a 13 MHz supply to the internal antenna generates a continuous low-power plasma. The high current beam pulses are generated by superimposing 50-70 kW from a pulsed 80 kW, 2 MHz amplifier. It is tuned for maximum H- beam current. The molybdenum converter is cesiated by the plasma heating cesium cartridges. A damping magnet suppresses the flow of coextracted electrons. The extracted beam of H- with current up to 60 mA, energy 65 keV is transported to the RFQ by electrostatic LEBT, comprised of two Einzel lenses. This RF SPS can work up to 99 days with beam current ~60 mA using ~30 mg of cesium. A similar internal antenna RF SPS was developed for the J-PARC synchrotron [31]. A schematic of this RF SPS is shown in Fig. 11. It has an internal SNS type antenna, conical cesiated collar for H- formation, extractor with permanent magnets, magnetic lenses, and magnetic correctors.

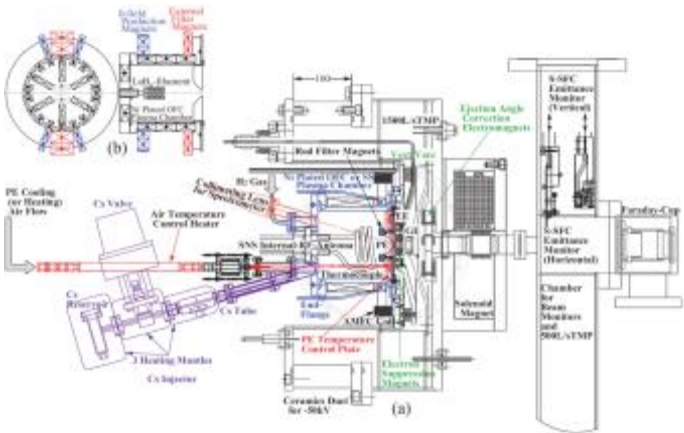

Fig. 11. Schematic of internal antenna RF SPS with cesiation in J-PARC, Japan.

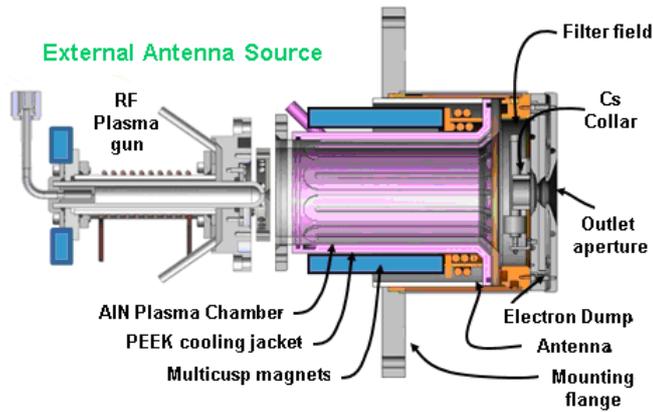

Fig. 12. Schematic of external antenna RF SPS with cesiation in SNS ORNL.

An external antenna RF SPS [32] was also developed for the SNS with the schematic shown in Fig. 12. In this RF SPS, the external antenna is located outside of the AlN ceramic discharge chamber and KEEP cooling jacket.

A similar external antenna RF SPS with cesiation was developed for Linac4 at CERN [33]. A schematic of this RF SPS is shown in Fig. 13. It includes an $Al_2O_3$ ceramic discharge chamber, external antenna with cusp magnets, conical cesiated converter and an extractor with permanent magnet and an electrostatic Einzel lens. 40 mA H- beam can be extracted from this RF SPS.

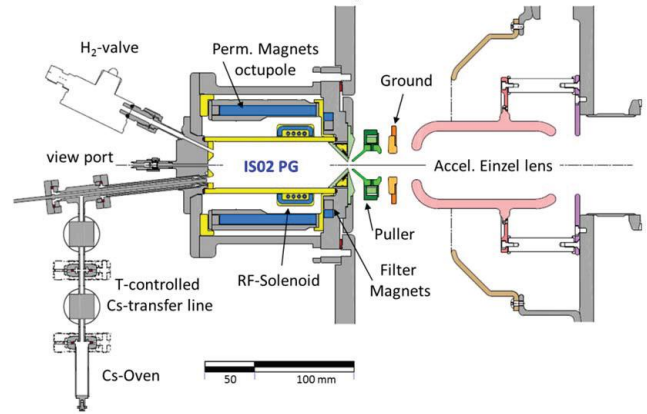

Fig. 13. Schematic of external antenna RF SPS with cesiation for Linac4 in CERN.

The Table below lists the parameters of Pulsed High Current H- Ion Sources for Accelerators. The most intense H- beam current is produced from a magnetron-planotron SPS with cesiation. The longest lifetime is produced by the RF SPS with cesiation at the ORNL SNS.

Table: Pulsed High Current H- Ion Sources for Accelerators

| H⁻ Source [reference] | Method | Discharge & Repetition Rate | Plasma & Beam Duty Factors | Average Beam Pulse Current | Extraction Aperture | Service Cycle/ Lifetime* | Extracted H⁻ Charge |
|---|---|---|---|---|---|---|---|
| BNL Operation | Magnetron Surface | 12-14A;130 V @ 7.5 Hz | 0.50 % 0.44 % | 110-120 mA | 2.8 mm ∅ | 6-8 months | 3.0 A·h |
| FNAL Operation | Magnetron Surface | 15 A;180 V @ 15 Hz | 0.345 % 0.3 % | 80 mA | 3.2 mm ∅ | 9 months | 1.6 A·h 3.2 A·h* |
| ISIS Operation | Penning Surface | 55 A; 70 V @ 50 Hz | 3.75 % 1.1 % | 55 mA | 0.6 x 10 mm² slit | 5 weeks* | 0.51 A·h* |
| CSNS Phase I | Penning Surface | ~50 A; ~100 V @ 25 Hz | 1.5 % 1.25 % | 50 mA | 0.6 x 10 mm² slit | 1 month* | 0.46 A·h* |
| INR RAS linac | Penning Surface | 100A;120V @ 50 Hz | 1 % 1 % | 20 mA | 1.0 x 10 mm² slit | Intermittent use | |
| LANSCE Operation | Filament driven converter | 30-35A; 180 V @ 120 Hz | 10 % 7.6 % | 16-18 mA | 9.8 mm ∅ | 4 weeks | 0.87 A·h |
| SNS Operation | Internal RF Antenna | CW 300 W 13 MHz & 60 Hz 60 kW 2 MHz | 6 % 5.94 % | >60 mA | 7 mm ∅ | 14 weeks | >7 A·h |
| J-PARC Operation | Internal RF antenna | CW 50 W 30 MHz & 25 Hz 22 kW 2 MHz | 2 % 1.25 % | 47 mA | 9 mm ∅ | 11 weeks | 1.1 A·h |
| CERN Linac4 | External RF antenna | 0.8 Hz 40 kW 2 MHz Pulsed H₂ | 0.07 % 0.05 % | 45 mA | 5.5 or 6.5 mm ∅ | 7 weeks | 0.026 A·h |

Neutral beam injector for ITER with Surface Plasma Source of Negative Ion with cesiation, invented by V. Dudnikov described in [ITER Neutral Beam Test Facility - 2017](ITER Neutral Beam Test Facility - 2017)

## VIII REEERENCES